\documentstyle[psfig,conf_iap,]{article}
\newcommand{\lya}{Ly-$\alpha$\ }
\newcommand{\dndz}{$dN/dz$}
\newcommand{\ggh}{\Gamma_{\rm HI}}
\newcommand{\nhi}{$N_{HI}$}
\newcommand{\nh}{N_{HI}}
\newcommand{\cm}{\rm cm} 
\newcommand{\magcir}{\ \raise -2.truept\hbox{\rlap{\hbox{$\sim$}}\raise5.truept
 	\hbox{$>$}\ }}		
\newcommand{\mincir}{\ \raise -2.truept\hbox{\rlap{\hbox{$\sim$}}\raise5.truept
	\hbox{$<$}\ }}			
\begin{document}
\heading{%
%Begin Heading
%
Exploring the Lyman Forest with VLT/UVES
%
%End Heading
} 
\par\medskip\noindent
\author{%
%Begin Author names
Stefano Cristiani$^{1,2}$, Simone Bianchi$^{3}$, Sandro
D'Odorico$^{3}$, Tae-Sun Kim$^{3}$ 
%End Author names
}
\address{
%First address
ST European Coordinating Facility, K.-Schwarzschild-Str. 2,
D-85748 Garching}
\address{
Department of Astronomy, Vicolo dell'Osservatorio 2, I-35122 Padova}
\address{
European Southern Observatory, K.-Schwarzschild-Str. 2, D-85748 Garching
}

\begin{abstract}
A sample of 8 QSOs with
$1.7 < z_{\rm em} < 3.7$ has been observed with VLT/UVES
at a typical resolution $45000$ and S/N $\sim
40-50$.
Thanks to the two-arm design of the spectrograph, a remarkable
efficiency has been achieved below 400nm and above 800nm, which
translates immediately in the possibility of obtaining new results,
especially at $z \mincir 2.5$. We report here new insight gained about
the evolution of the number density of \lya lines, their column
density distribution, the ionizing UV background and the cosmic baryon 
density.
\end{abstract}
\section{The number density evolution of \lya lines}
\begin{figure}[t]
\centerline{\vbox{
\psfig{figure=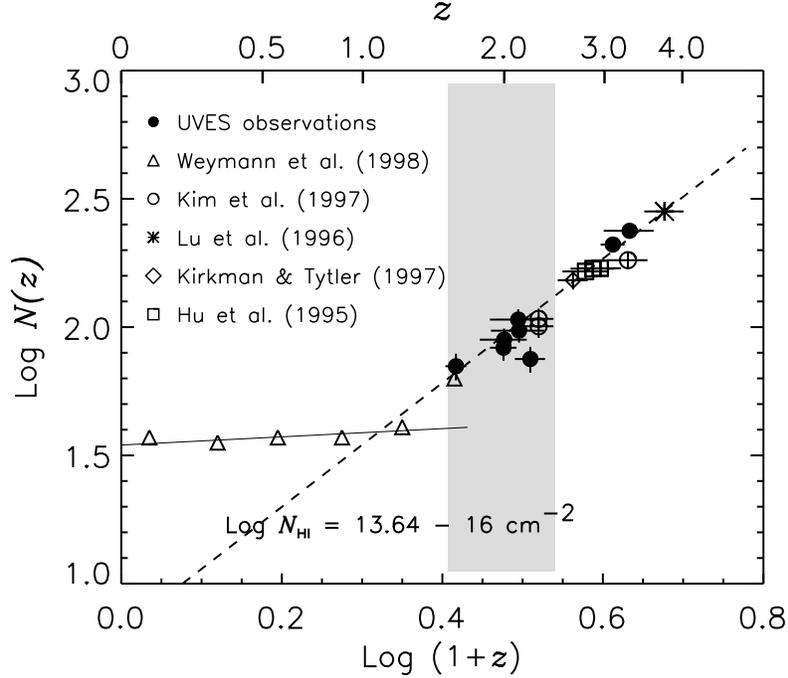,height=9.cm}
}}
\vskip -0.2cm
\caption[]{The number density evolution with $z$ \cite{kim02}
}
\label{dndz}
\end{figure}

The swift increase of the number of absorptions (and the average
opacity) with increasing
redshift is the most impressive property of the \lya forest.
Fig.~\ref{dndz} shows the number density evolution of the
\lya lines \cite{kim01,kim02} in the column density interval
\footnote{This range in \nhi\ has been chosen to allow a
comparison with the HST Key-Programme sample at $z < 1.5$ \cite{weymann98}
for which a threshold in equivalent width of 0.24 \AA\/ was adopted.}
$N_{HI} = 10^{13.64 - 16} \ {\rm cm}^{-2}$.
The long-dashed line is the maximum-likelihood fit 
to the data at $z > 1.5$ with the customary
parameterization: $N (z) = N_{0}
(1+z)^{\gamma} = (6.5 \pm 3.8)\,(1+z)^{2.4 \pm 0.2}$. 
The UVES \cite{dodorico00} observations imply that the turn-off in the
evolution does occur at $z \sim 1$, not at $z \sim 2$ as
previously suggested.  
The evolution of the $N(z)$ is governed by two main factors: the
Hubble expansion and the metagalactic UV background (UVB).
At high $z$ both the expansion, which decreases the density
and tends to increase the ionization,
and the UVB, which is increasing or non-decreasing 
with decreasing redshift, work in the same direction and
cause a steep evolution of the number of lines.
At low $z$, the UVB starts to decrease with decreasing
redshift, due to the reduced number and intensity of the ionizing sources, 
counteracting the Hubble expansion. As a result the evolution 
of the number of lines slows down.

Up to date numerical simulations \cite{theuns98} have been remarkably
successful in qualitatively reproducing the observed evolution,
however they predict the break in the \dndz\ power-law at a
redshift $z \sim 1.8$ that appears too high in the light of the new
UVES results. This suggests that the UVB implemented in the
simulations may not be the correct one: it was thought that at low
redshift QSOs are the main source of ionizing photons, and, since
their space density drops below $z\sim 2$, so does the UVB.
However, galaxies can produce a conspicuous ionizing flux too, perhaps
more significant than it was thought\cite{steidel01}. 
The galaxy contribution can keep the
UVB relatively high until at $z \sim 1$ the global star
formation rate in the Universe quickly decreases, determining the
qualitative change in the number density of lines.
Under relatively general assumptions, it is possible to
relate the observed number of lines above a given threshold in column
density or equivalent width to the
expansion, the UVB, the distribution in column density of 
the absorbers and the cosmology:
\begin{equation}
\left(dN \over dz\right)_{>N_{HI,\rm lim}} = 
C \left[(1+z)^5 \ggh^{-1}(z)\right]^{\beta-1} H^{-1}(z),
\label{eq:dndz}
\end{equation}
where $\ggh$ is the photoionization rate and the \nhi\ distribution  
is assumed to follow a power-law of index $\beta$, as discussed in the 
next section.
\begin{figure}[t]
\centerline{\vbox{
\psfig{figure=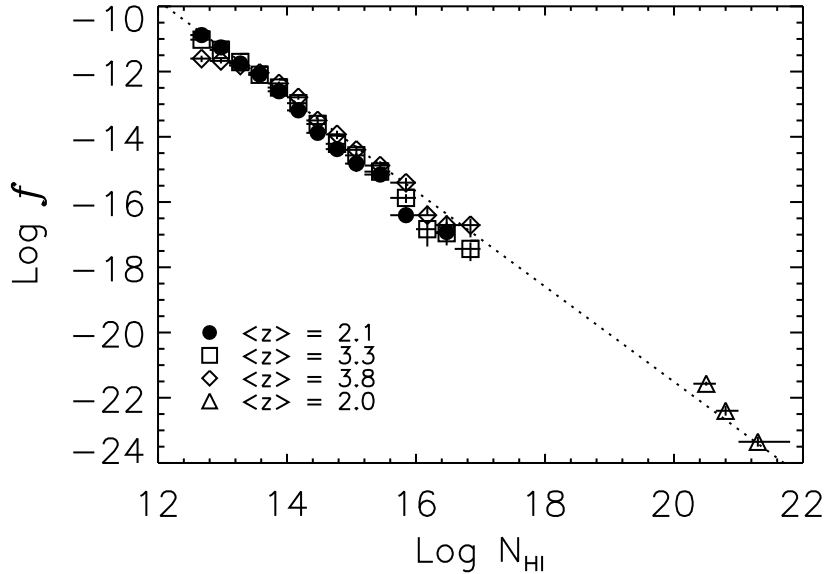,height=8.cm}
}}
\vskip -0.7cm
\caption[]{The column density distribution \cite{kim02}}
\label{NHI}
\end{figure}
\section{The column density distribution}
Fig.~\ref{NHI} shows the differential density distribution function
measured by UVES \cite{kim01,kim02}, that is the number of lines per
unit redshift path and per unit \nhi\ as a function of \nhi.  The
distribution basically follows a power-law $f( \nh ) \propto
\nh^{-1.5}$ extending over 10 orders of magnitude with little, but
significant deviations, which become more evident and easy to
interpret if the plot is transformed in the mass distribution of the
photoionized gas as a function of the density contrast, $\delta$,
\cite{schaye01}:
1) a flattening at $\log N_{HI}\mincir 13.5$ is partly due to line crowding
and partly to the turnover of the density distribution below the mean
density;
2) a steepening at $\log \nh \magcir 14$, with a deficiency of lines
that becomes more and more evident at lower z, reflects the fall-off
in the density distribution due to the onset of rapid, non-linear
collapse: the slope $\beta$ goes from about $-1.5$ at $<z> = 3.75$ to
$-1.7$ at $z < 2.4$ and recent HST STIS data \cite{dave01} confirm
that this trend continues at lower redshift
measuring at $z<0.3$ a slope of $-2.0$;
3) a flattening at $N_{HI} \magcir 10^{16}~\cm^{-2}$ can be attributed to
the flattening of the density distribution at $\delta \magcir 10^2$ due to
the virialization of collapsed matter.
Hydrodynamical simulations successfully reproduce this behaviour,
indicating that the derived matter distribution is indeed
consistent with what would be expected from gravitational instability.
\section{The ionizing background}
The last ingredient to be determined in Eq.~\ref{eq:dndz} is the
ionization rate. In a recent computation \cite{bianchi01} we have
investigated the contribution of galaxies to the UVB, exploring three
values for the fraction of ionizing photons that can escape the galaxy
ISM, $f_{esc} = 0.05, 0.1$ and $0.4$ (the latter value corresponds to
the Lyman-continuum flux detected by \cite{steidel01} in the composite
spectrum of 29 Lyman-break galaxies). 
Estimates of the UVB based on the proximity effect at high-$z$ and on
the ${\rm H}\alpha$ emission in high-latitude galactic clouds at
low-$z$ provide an upper limit on $f_{esc} \mincir 0.1$, consistent with
recent results on individual galaxies both at low-$z$
\cite{deharveng01,heckman01} and at $z\sim3$ \cite{giallongo01}.
Introducing a contribution of galaxies to the UVB,
the break in the \lya \dndz\ can be better
reproduced than with a pure QSO contribution \cite{bianchi01}. 
The agreement improves
considerably also at $z\magcir 3$. Besides, models with $\Omega_{\Lambda}
=0.7, \Omega_{M}=0.3$ describe the flat evolution of the absorbers
much better than $\Omega_{M}=1$.

A consistency check is provided by the evolution of the lower column
density lines. For $\log \nh \mincir 14$ the \nhi\ distribution is
flatter, and according to Eq.~\ref{eq:dndz} this translates directly
into a slower evolutionary rate, which is consistent
with the UVES observations\cite{kim01}: $\dndz_{(13.1<\nh<14)} \propto
(1+z)^{1.2\pm0.2}$. Another diagnostic can be
derived from the spectral shape of the UVB and its influence
on the intensity ratios of metal lines \cite{savaglio97}.
\section{The cosmic baryon density}
Given the cosmological scenario,
the amount of baryons required to produce the opacity of the
Lyman forest can be computed \cite{weinberg97} 
and a lower-bound to the cosmic baryon density derived
from the distribution of the \lya optical depths.
Applying this approach to the effective optical
depths measured in the UVES spectra, the estimated lower bound
$\Omega_b \magcir 0.010-0.016$
%%% from $z \sim 1.5$ to $4$
is consistent with the BBN value for a low D/H primordial abundance.
Most of the baryons reside in the Lyman
forest at $1.5<z<4$ with little change in the contribution to
$\Omega$ as a function of $z$.
Conversely, given the observed opacity, a higher UVB requires a
higher  $\Omega_b$. As pointed out by \cite{haehnelt01}, 
the large escape fraction measured by \cite{steidel01} would result
in an $\Omega_b \sim 0.06$ in strong conflict either with the D/H
or in general with the BBN or with the \lya opacity measurements.

\vfill
\end{document}